\documentclass[aps,prd,twocolumn,showpacs,groupedaddress,nofootinbib,longbibliography]{revtex4-1}
\expandafter\let\csname equation*\endcsname\relax 
\expandafter\let\csname endequation*\endcsname\relax 

\usepackage{stmaryrd}

\usepackage{times}
\usepackage{amsfonts}
\usepackage{graphicx}
\usepackage{amsmath}
\usepackage{amssymb}
\usepackage{color}
\usepackage{subfigure}
\usepackage{natbib}
\usepackage{tensor}
\usepackage{accents}
\usepackage[colorlinks=true, citecolor=blue]{hyperref}
\usepackage{xspace}

\def\beq{\begin{equation}}
\def\eeq{\end{equation}}
\def\bea{\begin{eqnarray}}
\def\eea{\end{eqnarray}}
\def\ben{\begin{enumerate}}
\def\een{\end{enumerate}}

\def\a{\alpha}
\def\b{\beta}
\def\g{\gamma}\def\G{\Gamma}

\def\e{\epsilon}

\def\l{\lambda}

\def\o{\omega}

\def\r{\rho}

\def\t{\tau}

\def\half{{\textstyle{\frac{1}{2}}}}

\def\w{\wedge}

\def\bE{{\bf E}}

\def\bB{{\bf B}}

\def\bj{{\bf j}}

\def\bv{{\bf v}}

\def\bs{{\bf s}}

\def\bv{{\bf v}}

\def\bs{\boldsymbol{\sigma}}

\begin{document}

\title{Structure of Aristotelian electrodynamics}

\author{Ted Jacobson}
\email{jacobson@umd.edu}

\affiliation{Kavli Institute for Theoretical Physics, University of California, Santa Barbara, CA 93106\\
Maryland Center for Fundamental Physics, 
University of Maryland, College Park, MD 20742}

\begin{abstract}
Aristotelian electrodynamics (AE) describes the regime of a plasma with a very strong
electric field that is not shorted out, with charge current determined completely 
by pair production and the balance of Lorentz 4-force against curvature radiation reaction.
Here it is shown how the principal null directions and associated eigenvalues of the field tensor 
govern AE, and how force-free electrodynamics arises smoothly from AE when the eigenvalues 
(and therefore the electric field in some frame) vanish. A criterion for validity of AE  
is proposed in terms of a pair of ``field curvature scalars" formed from the 
first derivative of the principal null directions.

\end{abstract}

\maketitle

\section{Introduction}

Gruzinov has proposed that force-free electrodynamics is inadequate for describing some relativistic magnetospheres, because of insufficient pair supply to short out the induced electric fields. 
In particular he argued that this is the case for ``weak pulsars" \cite{2013arXiv1303.4094G}. 
In this case, parts of the magnetosphere would 
fail to be force-free. When the fields are very strong and $\bE\cdot\bB$ is nonzero, charges will be accelerated to near the speed of light, reaching a steady state where the work done on them by the electric field is transferred to their radiation fields. In this situation, all charges move approximately at the speed of light, and the current is completely determined by the supply of charge. This regime was called 
{\it Aristotelian electrodynamics} (AE), because it is the velocity rather than the acceleration of the charges that is determined by the field. The name is quite fitting, since a central focus of Aristotelian mechanics was the regime of terminal velocity, in which the transients can be neglected \cite{2013arXiv1312.4057R}.
Consideration of this regime in pulsar electrodynamics was earlier 
discussed in \cite{1985MitAG..63..174H,1989A&A...225..479F}.

In the degenerate case when $\bE\cdot\bB=0$, if the field is magnetic ($B^2>E^2$), there exist Lorentz frames with vanishing electric field.  The charges therefore have no secular acceleration, so it can be a good approximation in strong, magnetic degenerate fields to neglect the energy and momentum of the charges altogether. In this {\it force-free} approximation, the current is entirely determined by the fields at one time (see e.g.\ \cite{2014MNRAS.445.2500G} for a review and references).
The Aristotelian and force-free approximations may coexist in one physical system that includes both
degenerate and non-degenerate regions. Both approximations 
result from neglecting of the mass of the charges in certain quantities. Together they comprise what Gruzinov called the {\it electrodynamics of massless charges} \cite{2012arXiv1205.3367G}.

Besides pulsars, another astrophysical setting where one would expect AE conditions to prevail is the magnetosphere of a compact binary system near coalescence. Even with copious pair creation, because of the motion of the binary, the fields change too rapidly for the charges to short out the electric field. 
One would therefore expect very strong charge acceleration to take place \cite{Sobacchi:2015yya}.
AE may also be relevant in a laboratory setting, for plasmas in Petawatt laser fields \cite{2014arXiv1404.4615G}.

In this paper I reformulate AE in a relativistically covariant form, and aim to elucidate the structure of the theory and its interface with force-free electrodynamics (FFE). The key observation is the central role played by the principal null directions of the electromagnetic field. 
Beyond that there is nothing new here.
The covariant form also applies in curved spacetime, so would be particularly useful in systems containing a black hole. The spacetime signature is $({+}{-}{-}{-})$ and the speed of light is sometimes set to $c=1$.

\section{Structure of electromagnetic fields}

A non-null electromagnetic field $F_{ab}$ at a point has two null eigenvectors $k_\pm$, with opposite eigenvalues,
\beq\label{eigen}
F^a{}_bk^b_\pm = \pm E_0\, k_\pm^a.
\eeq
The directions of these null eigenvectors are called the {\it principal null directions} (PNDs) 
of the field \cite{1986ssv..book.....P}.
The eigenvalue $\pm E_0$ is the electric field in a frame in which the electric and magnetic fields are parallel or one of them vanishes. 
In AE, $E_0\ne0$, and positive charges travel along one PND while negative charges travel along the other. In FFE, $E_0=0$, and the total current 4-vector also lies in the timelike plane spanned by the two PNDs. 
For null fields there is only one null eigenvector, with vanishing eigenvalue. 
 
A simple way to understand this structure is to observe that, at a given spacetime point,
any electromagnetic field 2-form $F$ can be presented in one of two canonical ways in terms of 
an adapted orthonormal set of 1-forms $\{dt, dx, dy, dz\}$:
\bea
F^{\rm generic} &=& E_0 \, dz\w dt + B_0\, dx\w dy,\label{generic}\\
F^{\rm null} &=& F_0 \, dz\w(dt-dx).\label{null}
\eea
The field in  \eqref{generic} has two 
null eigenvectors, $k_\pm = \partial_t\pm \partial_z$, with eigenvalues $\pm E_0$. 
The field in \eqref{null} has one null eigenvector, $\partial_t+ \partial_x$, 
with vanishing eigenvalue. 
The electric and magnetic fields in \eqref{generic} are parallel in the Lorentz frame  
defined by $\partial_t$. The general statement is that they are parallel in any frame lying in 
the timelike plane spanned by $k_+$ and $k_-$.
If the field is degenerate, i.e.\ if $F\w F=0$ ($\bE\cdot\bB=0$), then either $E_0=0$ or $B_0=0$; that is, either
$\bE$ or $\bB$ will vanish in these ``field eigenframes".
If the field is null then $\bE$ and $\bB$ are perpendicular and have the same magnitude in all frames.

The quantities $E_0$ and $B_0$ are related to Lorentz scalars by 
\bea
E_0^2 - B_0^2 &=& E^2 - B^2 = -\half F_{ab}F^{ab} \label{F2}\\
E_0B_0&=& \bE\cdot\bB = \tfrac18 \e^{abcd}F_{ab}F_{cd}. \label{E0B0}
\eea
To fix signs uniquely, we take $E_0\ge0$. 
$E_0^2$ and $B_0^2$ are given in terms of $a=E^2-B^2$ and $b=2\bE\cdot\bB$ by 
$(\sqrt{a^2 + b^2}\,\pm\, a)/2$, respectively. The PNDs can be specified by a pair of 
spatial unit vectors $\bv_\pm$ in a given frame
via $k_\pm^a\leftrightarrow (1,\bv_\pm)$.
Gruzinov gives an explicit expression for $\bv_\pm$
in terms of the electric and magnetic fields:
\beq\label{vpm}
\bv_\pm = \frac{\bE\times\bB \pm(E_0\bE + B_0\bB)}{E_0^2 + B^2}.
\eeq
[The denominator can also be written more symmetrically as $\half(E_0^2 +B_0^2 +E^2+ B^2)$.]

The PNDs can be constructed as quadratic expressions from the spinor factors (or eigenspinors) of the electromagnetic spinor \cite{1986ssv..book.....P}.\footnote{An electromagnetic field can be represented by the trace-free $2\times2$ matrix $\Phi=(\bE + i\bB)\cdot\bs$.
The Lorentz group acts by similarity transformation on $\Phi$, so the eigenvalues of $\Phi$ are Lorentz invariant. 
The properties of the Pauli matrices imply $\Phi^2 = (\bE + i\bB)\cdot(\bE + i\bB)\equiv(E_0+iB_0)^2I$,
so the eigenvalues are $\pm(E_0+iB_0)$. The eigenspinors $\lambda_\pm$ are the square roots of the PNDs of $F_{ab}$, in the sense that
$k_\pm^a \leftrightarrow (1, \lambda_\pm^\dagger\bs\lambda_\pm)$, with the normalization $\l_\pm^\dagger\l_\pm=1$.}
The Newman-Penrose formalism or other spinor techniques might therefore be useful for analytical and/or numerical studies of AE. 

\begin{figure}
\centering
\includegraphics[scale=.6]{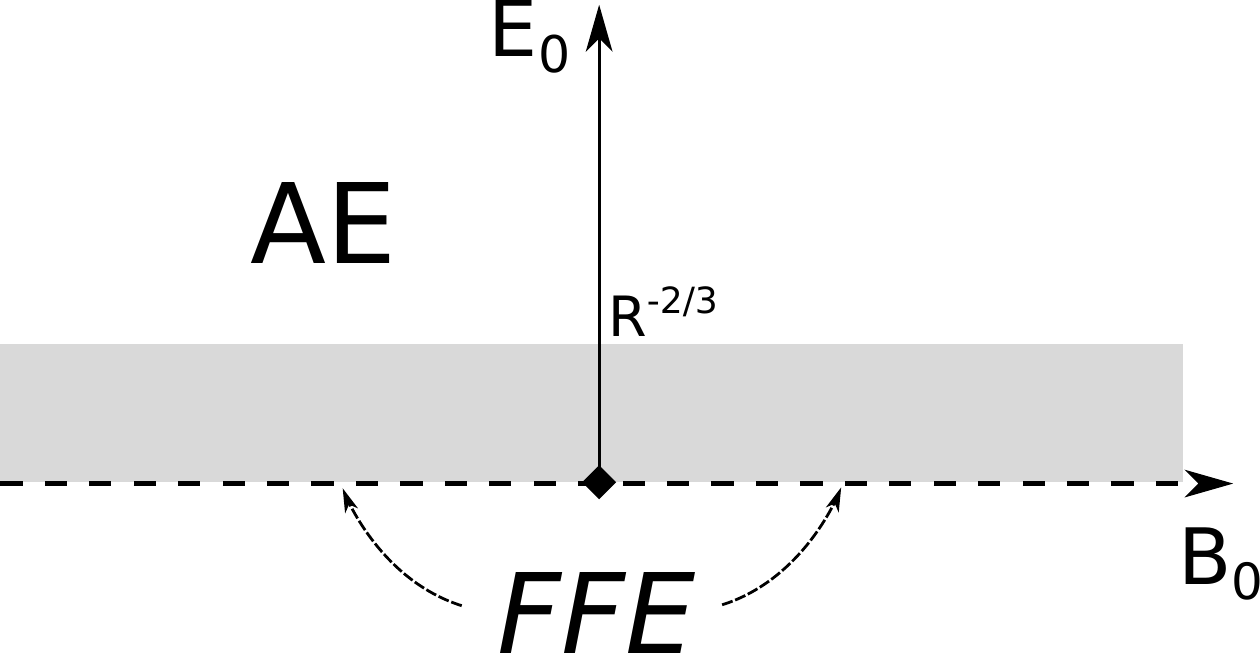}
\caption{The space of electromagnetic fields modulo Lorentz transformations.
The coordinates $(B_0,E_0)$ are the invariants defined in (\ref{F2},\ref{E0B0}). Degenerate fields are represented by the two axes. $(\pm B_0,0)$ are identified on the dashed lower boundary, so the space is actually a cone. AE applies above the grey strip of height $R^{-2/3}$, where $R$ is the curvature radius of the particle paths, 
reckoned in an instantaneous field eigenframe,
and units $e=mc^2=1$ are used. The dashed $B_0$ axis corresponds to magnetic or null ($B_0=0$) degenerate fields; FFE applies for these if the energy density of the field is much greater than that of the charges.}
\label{fieldplane}
\end{figure}

The set of fields modulo Lorentz transformations
can be represented as the upper half plane with vertical axis $E_0$ and horizontal axis $B_0$
(see Fig.~\ref{fieldplane}).
Degenerate fields have either $E_0=0$ or $B_0=0$, while null fields have $E_0=B_0=0$. 
When $E_0=0$,  the two values $\pm B_0$ 
label the same set of fields. The moduli space of fields is thus the cone obtained by identifying the positive-$B_0$ half of the dashed line in 
Fig.~\ref{fieldplane} with the negative half; 
the null fields lie at the vertex, and the magnetic and electric degenerate fields correspond to 
a pair of opposite rays on the cone. 

The current 4-vector lies in the plane spanned by the PNDs in both the AE and FFE regimes, though it is determined by different conditions. The distinguishing factor between these regimes is whether the PND eigenvalue is nonzero or zero. 
In the following we characterize these two regimes. 

The PND's define a pair of ``field curvature scalars",  
\beq\label{Rpm}
R_\pm^{-1} = \frac{|(k_\pm\cdot \nabla) k_\mp|}{|k_+\cdot k_-|/2}.
\eeq
These quantities have dimensions of inverse length and are 
invariant under arbitrary rescalings of $k_\pm$.\footnote{Since $k_-^2=0$, $(k_+\cdot\nabla)k_-$
is orthogonal to $k_-$, so $(k_+\cdot\nabla)k_-=Ak_- + v$ for some spacelike vector $v$ orthogonal to both $k_-$ and $k_+$, and  $|(k_+\cdot\nabla)k_-|=|v|$.
Under a rescaling $k_\pm \rightarrow \a_\pm k_\pm$ the coefficient $A$ is modified, and 
$v\rightarrow \a_+\a_-v$, so $|v|$ is simply multiplied by $|\a_+\a_-|$, which cancels against a similar factor from the denominator of \eqref{Rpm}.} 
The two radii  $R_\pm$ coincide if the PNDs are surface forming. 
For static electric or magnetic fields they coincide and are equal to the curvature radius of the field lines. 
More generally, they measure a rate at which one eigendirection bends away in the orthogonal spacelike direction when moving along the other PND. 
They are equal to twice the magnitudes of the spin coefficients $\pi$ and $\t$ associated with a null tetrad constructed from the PNDs \cite{1986ssv..book.....P}.
They may serve generally to define the relevant length scales determining the applicability of the AE approximation and the ``guiding center approximation" 
(see e.g.\ \cite{1987PhDT.......197B} and references therein).\footnote{I am grateful to Antony Speranza for suggesting that the invariants \eqref{Rpm} might supply the relevant notion of curvature length scale in this context.}

\section{Force-free electrodynamics}

If $E_0=0$, then the field is magnetic or null degenerate.
If the energy density of the field is much greater than that of the charges, 
then the energy and momentum of the charges can be neglected,
which means that the 4-force on the current can be set to zero, 
\beq\label{FF}
F_{ab}j^b=0.
\eeq
This condition in turn implies that the field is degenerate.
In the magnetic case, the FF current must be a linear combination of
the vectors $k_\pm^a$, since the latter span the kernel (i.e.\ the null eigenspace) of $F_{ab}$.
In the null case the PNDs coincide, the kernel of $F_{ab}$ is spanned by 
the unique PND and an orthogonal spacelike direction, and the current must be 
a linear combination of vectors in those directions.  
If there is only one sign of charge present, the current cannot be spacelike, 
so in the case of null fields it must be null.  

In either case, somewhat surprisingly, the current is determined by the fields at 
one time when \eqref{FF} and Maxwell's equations are imposed. 
One therefore has stand-alone evolution equations for magnetically dominated or null force-free fields,
without reference to the charge or current densities.
In the magnetic case the initial value problem for these equations is well-posed. 
 \cite{komissarov2002,palenzuela-etal2011,2013arXiv1307.7782P}.

\section{Aristotelian electrodynamics}

For any field with $E_0\ne0$, i.e.\ which is not magnetic or null degenerate, the electric field can accelerate charges without deflection along the spatial directions of the PNDs. In a strong field the charges accelerate until they radiate energy at the same rate as the Lorentz force supplies it to them---unless collisions with something slows
them down before that happens.\footnote{The charge acceleration could be halted, for example, by inverse Compton collisions with ambient photons, as in the stagnation surface scenario for black hole jets discussed in \cite{Broderick:2015swa}.} The power input is $\sim eE_0c$, 
reckoned in a field eigenframe in which the charge has speed $\approx c$, while
a charge moving with large Lorentz $\gamma$ factor on a path following a trajectory with radius of curvature $R$ emits curvature radiation with power $\frac23 e^2\g^4c/R^2$ \cite{1975ctf..book.....L}. Equating the input and output power yields $\g =(3E_0 R^2/2e)^{1/4}$.\footnote{The characteristic frequency of the radiation is $\o_c \sim \g^3c/R \sim (E_0/e)^{3/4} cR^{-1/2}$.} 
The 4-velocity of the charge is nearly parallel to the principal null vector $k_+^a$ if the charge is positive, and to $k_-^a$ if the charge is negative. 

When does AE apply? The electric field should be strong enough that it is a good approximation to treat the current of the charges as running parallel to the principal null directions. If the field is approximately static, it is presumably necessary that the 4-velocity is ``nearly lightlike" as reckoned in the approximately static frame,
and that the conditions for reaching terminal velocity have been met. 
Let us use units with $e=mc^2=1$ for a moment. The conditions just stated are that 
the terminal gamma factor must be relativistic, say $(E_0 R^2)^{1/4}>2$, and that the
particles reach that terminal velocity. This latter condition 
requires that the voltage drop  over a distance $\sim R$, reckoned in an instantaneous field eigenframe, be greater than the terminal gamma factor times the rest energy of the electron, $E_0 R > (E_0 R^2)^{1/4}$, i.e. $E_0 R^{2/3}>1$. The former condition is then automatically met in any relevant situation, since then
$R\gg R_e=1$ (the classical electron radius $e^2/mc^2$ in these units). This reasoning suggests that the condition of AE applicability is 
\beq\label{AEapp}
E_0> R^{-2/3}.
\eeq
 Note that  the range of applicability grows as $R$ grows (see Fig.~\ref{fieldplane}).
We conjecture that  the field curvature scalars \eqref{Rpm} provide the correct general interpretation of 
the length scale $R$ in \eqref{AEapp}.

Given a unit timelike vector $u^a$, we can normalize the principal null vectors by the condition $k^a_\pm u_a = 1$. Then, because the charge 4-velocities are parallel to $k_\pm^a$ in the AE regime, the current 
in a region where $E_0\ne 0$
takes the form
\beq\label{j}
j^a = \r_+ k_+^a - \r_-k_-^a,
\eeq
where $\r_+$ and $-\r_-$ are the densities of positive and negative charge in the frame $u^a$.
In terms of the velocities \eqref{vpm} relative to a fixed Lorentz frame, the 4-current density is given by 
\beq
 (\rho, \bj)=( \r_+ - \r_-,  \r_+ \bv_+ - \r_- \bv_-). 
 \eeq
Using the eigenvector property \eqref{eigen}, the 
Lorentz 4-force density acting on the current \eqref{j} is 
\beq
F^a{}_{b}j^b = E_0(\r_+ k_+^a + \r_-k_-^a).
\eeq
The power deposited into the charges in the frame $u^a$ is $E_0(\r_+ + \r_-)$.

The charge density is determined by $F_{ab}$ without time derivatives via Gauss' law,
$\r=\nabla\cdot\bE$.  Since the current
4-vector $j^a$ \eqref{j} lies in the plane spanned by the two null eigenvectors, the remaining freedom 
in $j^a$ is only one function. If all the charges in a given region have the same sign, then the current is 
null and thus fully determined. If instead both signs of charge are present (nonzero pair multiplicity),
the charge densities are determined by pair production and subsequent propagation subject to 
the continuity equations,
\beq\label{cons}
\nabla_a (\r_+ k_+^a) = \nabla_a (\r_- k_-^a) =\Gamma.
\eeq
The pair creation rate $\G$ depends on $E_0$ and the photon density. 
Once the form of $\G$ is specified, Maxwell's equations, together with Eqs.\ \eqref{j} and \eqref{cons} and initial values for $\r_\pm$,
determine the time derivatives in terms of the field and charge densities at a given time.
This system of equations is thus naively deterministic. (Whether it defines a well-posed initial value problem has not been examined, although Gruzinov has evolved them numerically and the solutions seem to behave well \cite{2013arXiv1303.4094G,2015arXiv150305158G}.)

\subsection{Example: Gruzinov's device}

To illustrate the workings of AE/FFE, Gruzinov considered 
an arrangement in two spatial dimensions where FFE and AE regimes coexist side by side.
He called this the ``Device" \cite{2014arXiv1402.1520G} (see Fig.\ 2 in this reference for 
an illustration). 
In the Device,
the $y=0$ plane is a conductor, and
opposing FFE Poynting fluxes propagating in the $\pm x$ directions
collide in an AE zone where the energy is
converted to curvature radiation. The charges are electrically attracted toward  
the image charges and repelled by the image currents below the conducting plane.
In the FFE zone the charges move rectilinearly at the speed of light, so the electric and magnetic forces
balance. This zone transitions to an AE radiation zone, in analogy with the transition 
outside a weak pulsar, without any discontinuity. In the AE zone, the magnetic field is weaker than
the electric one, so the net force attracts the charges to the conductor. 
Here I use the stationary Device to illustrate 
the formulation given above.

The stationary field can be expressed as 
\beq
F = E(y)dy\w[dt+ \b(x) dx],
\eeq
corresponding to an electric field $E$ in the $y$ direction and a magnetic field $-\b E$ 
in the $z$ direction. $F$ is a simple 2-form (as is any electromagnetic field in 2+1 dimensions), 
so it is degenerate. That is, one or both of $E_0$ and $B_0$ vanish.

For $\beta^2 <1$ the field is electric, so it is $B_0$ that vanishes. 
There are two PNDs,
\beq\label{pnd1}
k_\pm = \partial_t -\beta\, \partial_x \pm \sqrt{1-\b^2}\, \partial_y,
\eeq
with eigenvalues
\beq\label{E0}
E_0 = \pm E\sqrt{1-\b^2}.
\eeq
The frames  in which the magnetic field vanishes are those in the $k_+k_-$ plane. From \eqref{pnd1} we see that these have $x$-coordinate velocity $-\b$ and any $y$-coordinate velocity with magnitude less than or equal to $\sqrt{1-\b^2}$. A positive charge moving in the direction of the null vector $k_+$ 
feels a Lorentz force proportional to $k_+$. If $\b=0$ the spatial direction of such motion is just that of the electric field, $\partial_y$, while if $\b$ is nonzero there is also a $\partial_x$ component. 

For $\b^2=1$ the field is null, and the two PNDs coalesce into a single PND with vanishing eigenvalue.
In this case the field takes the form $F=E(y) dy\w(dt\pm dx)$, which is a solution of the force-free equations. 
In fact it has the form of the simple class of null Poynting flux solutions discussed in \cite{2014MNRAS.445.2500G}.
For $\b^2 >1$ the field is magnetic,  but it is not a force-free solution for any nontrivial choice of the functions $E$ and $\b$, so that case plays no role here.

Let us now consider the AE field equations in the electric case $\b^2<1$. First, the field 
satisfies Faraday's law, $dF=0$, by inspection. 
The remaining AE equations require that (i) the current computed from $F$ according to Maxwell's equations is equal to \eqref{j} and (ii) the continuity equation \eqref{cons} holds.  Assuming there is no pair creation ($\G=0$), 
and that only electrons are present, the continuity equation follows from the Maxwell equation, 
$\partial_b F^{ab}= -\r_-k_-^a$. This equation implies $\r_-=-E_{,y}$, and $E_{,y}/E = -\b_{,x}/\sqrt{1-\b^2}$.
Since $E$ depends only on $y$, and $\b$ depends only on $x$, the general solution is given by
\beq
E=E(0)\exp(-y/\ell),\qquad \b=\sin (x/\ell),
\eeq
where $\ell$ is a constant length. This solution matches smoothly to
the force-free Poynting flux solution with $\b=\pm1$ at $x/\ell=\pm \pi/2$. 
The charges and Poynting flux are thus flowing inwards toward $x=0$ from both sides.

What about the validity of the approximations for the Device? In the FF zone the 
charges move in straight lines at the speed of light, which can presumably be understood as an ultrarelativistic approximation.
Where the radiation zone begins, $E_0$ starts out at zero \eqref{E0}. 
According to \eqref{AEapp}, AE therefore
applies at the transition only if $R=\infty$, where $R$ is the radius of curvature in an 
instantaneous field eigenframe.  
We proposed that the invariant $R_-$ defined in \eqref{Rpm} may capture the relevant curvature radius. 
The vector fields \eqref{pnd1} are surface forming, so $R_+=R_-$, and we find 
$R_\pm^{-1}=\partial_x(1-\b^2)^{-1/2}$. This diverges at $x/\ell=\pi/2$, so $R_\pm\rightarrow 0$ there,
hence for any $E(0)$ there is a region close to $x/\ell=\pi/2$ where the applicability of AE is questionable. Specifically, using \eqref{E0}, the condition \eqref{AEapp} becomes $E(0)e^{-y/\ell}> \ell^{-2/3}[\sin(x/\ell)/\cos^2(x/\ell)]^{2/3}$. 

\section{Summary}

Aristotelian Electrodynamics \cite{2013arXiv1303.4094G} is the theory of a plasma with very strong electric
field, whose charges move ultrarelativistically, essentially at the speed of 
light in the direction of the local electric field.
In this paper we have formulated AE 
in a way that brings out 
the central role played by the principal null 
directions $k^a_\pm$ of the electromagnetic field tensor \eqref{eigen}. 
These determine the current \eqref{j} up to the densities of positive and negative charge, and can be used to construct the field curvature scalars \eqref{Rpm} relevant to applicability of the AE approximation.
The total charge density is determined by the divergence of the electric field, so just one function in the 4-current, say the positive charge density, remains undetermined by the field. 
This is determined by the pair production rate $\Gamma$  and the continuity equation \eqref{cons}.
The rate of 4-momentum density deposited into the charges is proportional to the eigenvalue $E_0$ \eqref{eigen}, and this eigenvalue can also be used to parametrize $\Gamma$.

\begin{acknowledgments}
I am grateful to Sam Gralla and Antony Speranza for helpful comments on a draft, 
to Antony for extensive discussions about the field curvature scalars, and to Carlo Rovelli for instruction
on Aristotelian physics.
This research was supported in part by the 
National Science Foundation under grants No. PHY-1407744, and PHY11-25915.
\end{acknowledgments}

\bibliography{AE}

\begin{thebibliography}{17}%
\makeatletter
\providecommand \@ifxundefined [1]{%
 \@ifx{#1\undefined}
}%
\providecommand \@ifnum [1]{%
 \ifnum #1\expandafter \@firstoftwo
 \else \expandafter \@secondoftwo
 \fi
}%
\providecommand \@ifx [1]{%
 \ifx #1\expandafter \@firstoftwo
 \else \expandafter \@secondoftwo
 \fi
}%
\providecommand \natexlab [1]{#1}%
\providecommand \enquote  [1]{``#1''}%
\providecommand \bibnamefont  [1]{#1}%
\providecommand \bibfnamefont [1]{#1}%
\providecommand \citenamefont [1]{#1}%
\providecommand \href@noop [0]{\@secondoftwo}%
\providecommand \href [0]{\begingroup \@sanitize@url \@href}%
\providecommand \@href[1]{\@@startlink{#1}\@@href}%
\providecommand \@@href[1]{\endgroup#1\@@endlink}%
\providecommand \@sanitize@url [0]{\catcode `\\12\catcode `\$12\catcode
  `\&12\catcode `\#12\catcode `\^12\catcode `\_12\catcode `\%12\relax}%
\providecommand \@@startlink[1]{}%
\providecommand \@@endlink[0]{}%
\providecommand \url  [0]{\begingroup\@sanitize@url \@url }%
\providecommand \@url [1]{\endgroup\@href {#1}{\urlprefix }}%
\providecommand \urlprefix  [0]{URL }%
\providecommand \Eprint [0]{\href }%
\providecommand \doibase [0]{http://dx.doi.org/}%
\providecommand \selectlanguage [0]{\@gobble}%
\providecommand \bibinfo  [0]{\@secondoftwo}%
\providecommand \bibfield  [0]{\@secondoftwo}%
\providecommand \translation [1]{[#1]}%
\providecommand \BibitemOpen [0]{}%
\providecommand \bibitemStop [0]{}%
\providecommand \bibitemNoStop [0]{.\EOS\space}%
\providecommand \EOS [0]{\spacefactor3000\relax}%
\providecommand \BibitemShut  [1]{\csname bibitem#1\endcsname}%
\let\auto@bib@innerbib\@empty
\bibitem [{\citenamefont {{Gruzinov}}(2013)}]{2013arXiv1303.4094G}%
  \BibitemOpen
  \bibfield  {author} {\bibinfo {author} {\bibfnamefont {A.}~\bibnamefont
  {{Gruzinov}}},\ }\bibfield  {title} {\enquote {\bibinfo {title}
  {{Aristotelian Electrodynamics solves the Pulsar: Lower Efficiency of Strong
  Pulsars}},}\ }\href@noop {} {\bibfield  {journal} {\bibinfo  {journal} {ArXiv
  e-prints}\ } (\bibinfo {year} {2013})},\ \Eprint
  {http://arxiv.org/abs/1303.4094} {arXiv:1303.4094 [astro-ph.HE]} \BibitemShut
  {NoStop}%
\bibitem [{\citenamefont {{Rovelli}}(2013)}]{2013arXiv1312.4057R}%
  \BibitemOpen
  \bibfield  {author} {\bibinfo {author} {\bibfnamefont {C.}~\bibnamefont
  {{Rovelli}}},\ }\bibfield  {title} {\enquote {\bibinfo {title} {{Aristotle's
  Physics: a Physicist's Look}},}\ }\href@noop {} {\bibfield  {journal}
  {\bibinfo  {journal} {ArXiv e-prints}\ } (\bibinfo {year} {2013})},\ \Eprint
  {http://arxiv.org/abs/1312.4057} {arXiv:1312.4057} \BibitemShut {NoStop}%
\bibitem [{\citenamefont {{Herold}}\ \emph {et~al.}(1985)\citenamefont
  {{Herold}}, \citenamefont {{Ertl}},\ and\ \citenamefont
  {{Ruder}}}]{1985MitAG..63..174H}%
  \BibitemOpen
  \bibfield  {author} {\bibinfo {author} {\bibfnamefont {H.}~\bibnamefont
  {{Herold}}}, \bibinfo {author} {\bibfnamefont {T.}~\bibnamefont {{Ertl}}}, \
  and\ \bibinfo {author} {\bibfnamefont {H.}~\bibnamefont {{Ruder}}},\
  }\bibfield  {title} {\enquote {\bibinfo {title} {{Generation of relativistic
  particles in pulsar magnetospheres}},}\ }\href@noop {} {\bibfield  {journal}
  {\bibinfo  {journal} {Mitteilungen der Astronomischen Gesellschaft Hamburg}\
  }\textbf {\bibinfo {volume} {63}},\ \bibinfo {pages} {174} (\bibinfo {year}
  {1985})}\BibitemShut {NoStop}%
\bibitem [{\citenamefont {{Finkbeiner}}\ \emph {et~al.}(1989)\citenamefont
  {{Finkbeiner}}, \citenamefont {{Herold}}, \citenamefont {{Ertl}},\ and\
  \citenamefont {{Ruder}}}]{1989A&A...225..479F}%
  \BibitemOpen
  \bibfield  {author} {\bibinfo {author} {\bibfnamefont {B.}~\bibnamefont
  {{Finkbeiner}}}, \bibinfo {author} {\bibfnamefont {H.}~\bibnamefont
  {{Herold}}}, \bibinfo {author} {\bibfnamefont {T.}~\bibnamefont {{Ertl}}}, \
  and\ \bibinfo {author} {\bibfnamefont {H.}~\bibnamefont {{Ruder}}},\
  }\bibfield  {title} {\enquote {\bibinfo {title} {{Effects of radiation
  damping on particle motion in pulsar vacuum fields}},}\ }\href@noop {}
  {\bibfield  {journal} {\bibinfo  {journal} {Astronomy and Astrophysics}\
  }\textbf {\bibinfo {volume} {225}},\ \bibinfo {pages} {479--487} (\bibinfo
  {year} {1989})}\BibitemShut {NoStop}%
\bibitem [{\citenamefont {{Gralla}}\ and\ \citenamefont
  {{Jacobson}}(2014)}]{2014MNRAS.445.2500G}%
  \BibitemOpen
  \bibfield  {author} {\bibinfo {author} {\bibfnamefont {S.~E.}\ \bibnamefont
  {{Gralla}}}\ and\ \bibinfo {author} {\bibfnamefont {T.}~\bibnamefont
  {{Jacobson}}},\ }\bibfield  {title} {\enquote {\bibinfo {title} {{Spacetime
  approach to force-free magnetospheres}},}\ }\href {\doibase
  10.1093/mnras/stu1690} {\bibfield  {journal} {\bibinfo  {journal}
  {Mon.~Not.~Roy.~Astron.~Soc.}\ }\textbf {\bibinfo {volume} {445}},\ \bibinfo
  {pages} {2500--2534} (\bibinfo {year} {2014})},\ \Eprint
  {http://arxiv.org/abs/1401.6159} {arXiv:1401.6159 [astro-ph.HE]} \BibitemShut
  {NoStop}%
\bibitem [{\citenamefont {{Gruzinov}}(2012)}]{2012arXiv1205.3367G}%
  \BibitemOpen
  \bibfield  {author} {\bibinfo {author} {\bibfnamefont {A.}~\bibnamefont
  {{Gruzinov}}},\ }\bibfield  {title} {\enquote {\bibinfo {title}
  {{Electrodynamics of Massless Charges with Application to Pulsars}},}\
  }\href@noop {} {\bibfield  {journal} {\bibinfo  {journal} {ArXiv e-prints}\ }
  (\bibinfo {year} {2012})},\ \Eprint {http://arxiv.org/abs/1205.3367}
  {arXiv:1205.3367 [astro-ph.HE]} \BibitemShut {NoStop}%
\bibitem [{\citenamefont {Sobacchi}\ and\ \citenamefont
  {Vietri}(2015)}]{Sobacchi:2015yya}%
  \BibitemOpen
  \bibfield  {author} {\bibinfo {author} {\bibfnamefont {Emanuele}\
  \bibnamefont {Sobacchi}}\ and\ \bibinfo {author} {\bibfnamefont {Mario}\
  \bibnamefont {Vietri}},\ }\bibfield  {title} {\enquote {\bibinfo {title}
  {{Some comments on the electrodynamics of binary pulsars}},}\ }\href
  {\doibase 10.1093/mnras/stv718} {\bibfield  {journal} {\bibinfo  {journal}
  {Mon.Not.Roy.Astron.Soc.}\ }\textbf {\bibinfo {volume} {450}},\ \bibinfo
  {pages} {2116--2121} (\bibinfo {year} {2015})},\ \Eprint
  {http://arxiv.org/abs/1503.08014} {arXiv:1503.08014 [astro-ph.HE]}
  \BibitemShut {NoStop}%
\bibitem [{\citenamefont
  {{Gruzinov}}(2014{\natexlab{a}})}]{2014arXiv1404.4615G}%
  \BibitemOpen
  \bibfield  {author} {\bibinfo {author} {\bibfnamefont {A.}~\bibnamefont
  {{Gruzinov}}},\ }\bibfield  {title} {\enquote {\bibinfo {title} {{Laboratory
  gamma-ray pulsar}},}\ }\href@noop {} {\bibfield  {journal} {\bibinfo
  {journal} {ArXiv e-prints}\ } (\bibinfo {year} {2014}{\natexlab{a}})},\
  \Eprint {http://arxiv.org/abs/1404.4615} {arXiv:1404.4615 [astro-ph.HE]}
  \BibitemShut {NoStop}%
\bibitem [{\citenamefont {{Penrose}}\ and\ \citenamefont
  {{Rindler}}(1986)}]{1986ssv..book.....P}%
  \BibitemOpen
  \bibfield  {author} {\bibinfo {author} {\bibfnamefont {R.}~\bibnamefont
  {{Penrose}}}\ and\ \bibinfo {author} {\bibfnamefont {W.}~\bibnamefont
  {{Rindler}}},\ }\href@noop {} {\emph {\bibinfo {title} {{Spinors and
  space-time. Volume 2: Spinor and twistor methods in space-time geometry}}}}\
  (\bibinfo  {publisher} {Cambridge University Press},\ \bibinfo {year}
  {1986})\BibitemShut {NoStop}%
\bibitem [{\citenamefont {{Boghosian}}(1987)}]{1987PhDT.......197B}%
  \BibitemOpen
  \bibfield  {author} {\bibinfo {author} {\bibfnamefont {B.~M.}\ \bibnamefont
  {{Boghosian}}},\ }\emph {\bibinfo {title} {{Covariant Lagrangian Methods of
  Relativistic Plasma Theory}}},\ \href@noop {} {Ph.D. thesis},\ \bibinfo
  {school} {UNIVERSITY OF CALIFORNIA, DAVIS.} (\bibinfo {year} {1987}),\
  \Eprint {http://arxiv.org/abs/physics/0307148} {arXiv:physics/0307148}
  \BibitemShut {NoStop}%
\bibitem [{\citenamefont {{Komissarov}}(2002)}]{komissarov2002}%
  \BibitemOpen
  \bibfield  {author} {\bibinfo {author} {\bibfnamefont {S.~S.}\ \bibnamefont
  {{Komissarov}}},\ }\bibfield  {title} {\enquote {\bibinfo {title}
  {{Time-dependent, force-free, degenerate electrodynamics}},}\ }\href
  {\doibase 10.1046/j.1365-8711.2002.05313.x} {\bibfield  {journal} {\bibinfo
  {journal} {Mon.~Not.~Roy.~Astron.~Soc.}\ }\textbf {\bibinfo {volume} {336}},\
  \bibinfo {pages} {759--766} (\bibinfo {year} {2002})},\ \Eprint
  {http://arxiv.org/abs/astro-ph/0202447} {astro-ph/0202447} \BibitemShut
  {NoStop}%
\bibitem [{\citenamefont {{Palenzuela}}\ \emph {et~al.}(2011)\citenamefont
  {{Palenzuela}}, \citenamefont {{Bona}}, \citenamefont {{Lehner}},\ and\
  \citenamefont {{Reula}}}]{palenzuela-etal2011}%
  \BibitemOpen
  \bibfield  {author} {\bibinfo {author} {\bibfnamefont {C.}~\bibnamefont
  {{Palenzuela}}}, \bibinfo {author} {\bibfnamefont {C.}~\bibnamefont
  {{Bona}}}, \bibinfo {author} {\bibfnamefont {L.}~\bibnamefont {{Lehner}}}, \
  and\ \bibinfo {author} {\bibfnamefont {O.}~\bibnamefont {{Reula}}},\
  }\bibfield  {title} {\enquote {\bibinfo {title} {{Robustness of the
  Blandford-Znajek mechanism}},}\ }\href {\doibase
  10.1088/0264-9381/28/13/134007} {\bibfield  {journal} {\bibinfo  {journal}
  {Classical and Quantum Gravity}\ }\textbf {\bibinfo {volume} {28}},\ \bibinfo
  {eid} {134007} (\bibinfo {year} {2011})},\ \Eprint
  {http://arxiv.org/abs/1102.3663} {arXiv:1102.3663 [astro-ph.HE]} \BibitemShut
  {NoStop}%
\bibitem [{\citenamefont {{Pfeiffer}}\ and\ \citenamefont
  {{MacFadyen}}(2013)}]{2013arXiv1307.7782P}%
  \BibitemOpen
  \bibfield  {author} {\bibinfo {author} {\bibfnamefont {H.~P.}\ \bibnamefont
  {{Pfeiffer}}}\ and\ \bibinfo {author} {\bibfnamefont {A.~I.}\ \bibnamefont
  {{MacFadyen}}},\ }\bibfield  {title} {\enquote {\bibinfo {title}
  {{Hyperbolicity of Force-Free Electrodynamics}},}\ }\href@noop {} {\bibfield
  {journal} {\bibinfo  {journal} {ArXiv e-prints}\ } (\bibinfo {year}
  {2013})},\ \Eprint {http://arxiv.org/abs/1307.7782} {arXiv:1307.7782 [gr-qc]}
  \BibitemShut {NoStop}%
\bibitem [{\citenamefont {Broderick}\ and\ \citenamefont
  {Tchekhovskoy}(2015)}]{Broderick:2015swa}%
  \BibitemOpen
  \bibfield  {author} {\bibinfo {author} {\bibfnamefont {A.~E.}\ \bibnamefont
  {Broderick}}\ and\ \bibinfo {author} {\bibfnamefont {A.}~\bibnamefont
  {Tchekhovskoy}},\ }\bibfield  {title} {\enquote {\bibinfo {title}
  {{Horizon-Scale Lepton Acceleration in Jets: Explaining the Compact Radio
  Emission in M87}},}\ }\href@noop {} {\  (\bibinfo {year} {2015})},\ \Eprint
  {http://arxiv.org/abs/1506.04754} {arXiv:1506.04754 [astro-ph.HE]}
  \BibitemShut {NoStop}%
\bibitem [{\citenamefont {{Landau}}\ and\ \citenamefont
  {{Lifshitz}}(1975)}]{1975ctf..book.....L}%
  \BibitemOpen
  \bibfield  {author} {\bibinfo {author} {\bibfnamefont {L.~D.}\ \bibnamefont
  {{Landau}}}\ and\ \bibinfo {author} {\bibfnamefont {E.~M.}\ \bibnamefont
  {{Lifshitz}}},\ }\href@noop {} {\emph {\bibinfo {title} {{The classical
  theory of fields}}}}\ (\bibinfo  {publisher} {Pergamon Press},\ \bibinfo
  {year} {1975})\BibitemShut {NoStop}%
\bibitem [{\citenamefont {{Gruzinov}}(2015)}]{2015arXiv150305158G}%
  \BibitemOpen
  \bibfield  {author} {\bibinfo {author} {\bibfnamefont {A.}~\bibnamefont
  {{Gruzinov}}},\ }\bibfield  {title} {\enquote {\bibinfo {title} {{No-Hair
  Theorem for Weak Pulsar}},}\ }\href@noop {} {\bibfield  {journal} {\bibinfo
  {journal} {ArXiv e-prints}\ } (\bibinfo {year} {2015})},\ \Eprint
  {http://arxiv.org/abs/1503.05158} {arXiv:1503.05158 [astro-ph.HE]}
  \BibitemShut {NoStop}%
\bibitem [{\citenamefont
  {{Gruzinov}}(2014{\natexlab{b}})}]{2014arXiv1402.1520G}%
  \BibitemOpen
  \bibfield  {author} {\bibinfo {author} {\bibfnamefont {A.}~\bibnamefont
  {{Gruzinov}}},\ }\bibfield  {title} {\enquote {\bibinfo {title} {{How Pulsars
  Shine: Poynting Flux Annihilation}},}\ }\href@noop {} {\bibfield  {journal}
  {\bibinfo  {journal} {ArXiv e-prints}\ } (\bibinfo {year}
  {2014}{\natexlab{b}})},\ \Eprint {http://arxiv.org/abs/1402.1520}
  {arXiv:1402.1520 [astro-ph.HE]} \BibitemShut {NoStop}%
\end{thebibliography}%

\end{document}